\documentclass{cjpsuppl}
 \usepackage{epsf,epsfig} 
\newcommand{\nn}{\nonumber}

\def\xb{\overline{x}}

\def\veps{\varepsilon}
\def\als{\alpha_s}
\def\vk{{\bf k}_{\perp}}

\def\gev{\,{\rm GeV}}
 \begin{document}
 \slacs{.6mm}
 \title{Deeply virtual electro-production of vector mesons and spin effects}
 \authori{S.V.Goloskokov}
 \addressi{ Bogoliubov Laboratory of Theoretical  Physics,
  Joint Institute for Nuclear Research, Dubna, Russia}
 \authorii{}    \addressii{}
 \authoriii{}   \addressiii{}
 \authoriv{}    \addressiv{}
 \authorv{}     \addressv{}
 \authorvi{}    \addressvi{}
 \headtitle{ Deeply virtual electro-production of vector mesons\ldots}
 \headauthor{S.\,V. Goloskokov}
 \lastevenhead{S.\,V. Goloskokov: Deeply virtual electro-production of vector mesons \ldots}
 \refnum{}
 \daterec{} 
 \suppl{A}  \year{2006} \setcounter{page}{1}
 \maketitle

 \begin{abstract}
We study light vector meson photoproduction at small $x$ on the
basis of the generalized parton distribution (GPD). Our results on
the cross section and spin density matrix elements (SDME) are in
fair agreement with DESY experiments.
 \end{abstract}

 \section{Introduction}

Vector meson leptoproduction at large energies  \cite{gk05} is one
of the important processes which is sensitive to the GPDs. At
small $x$-Bjorken ($x$) the process factorizes \cite{fact} into a
hard meson leptoproduction off gluons and GPD (Fig.1).
\begin{figure}
\centering \mbox{\epsfysize=30mm\epsffile{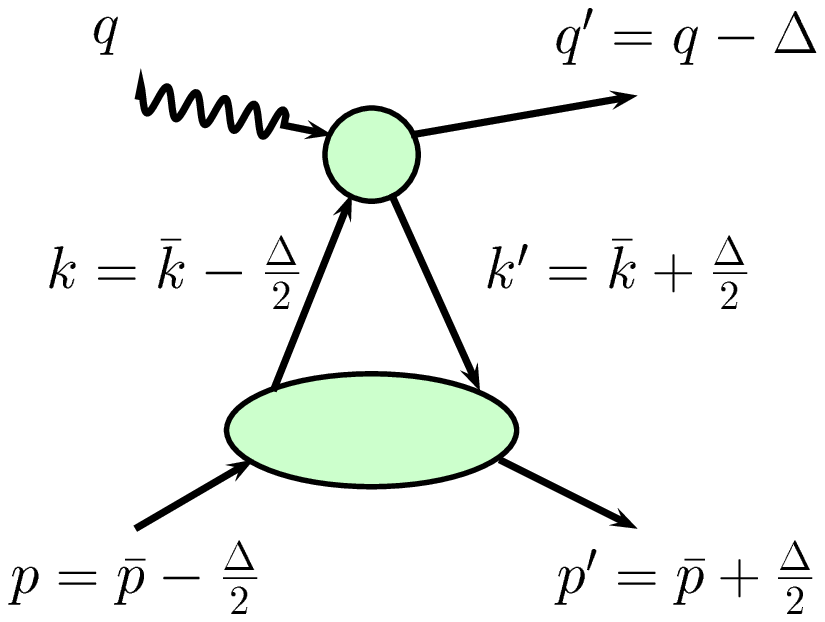}}
\caption{The handbag diagram for the meson electroproduction off
proton.} \label{kt_h}
\end{figure}

The generally used standard collinear approximation (SCA) leads to
some essential problems.  The cross section for longitudinally
polarized virtual photons which  predominates for the large photon
virtualities exceeds the data by a large factor if calculated in
the SCA \cite{mpw}. Only the amplitude with a longitudinally
polarized virtual photon and vector meson (LL) amplitude can be
calculated in the SCA. The transition amplitudes for transversally
polarized photons $\gamma _{\perp }^{\ast }\rightarrow V_{\perp }$
(TT) and $\gamma _{\perp }^{\ast }\rightarrow V_{L}$ (TL) exhibit
in the SCA the infrared singularities \cite{mp} which  breakdown
 factorization.

Here we use the modified perturbative approach (MPA)
\cite{sterman} which includes the quark transverse degrees of
freedom accompanied by Sudakov suppressions. In the MPA one can
solve all mentioned problems. Transverse quark degrees of freedom
and Sudakov factors suppress the contribution from the end-point
region in the LL amplitude. The transverse quark momentum
regularizes the end-point singularities in TT and TL amplitudes.
This leads to reasonable agreement of our results with H1 and ZEUS
data \cite{h1,zeus}  for electroproduced $\rho$ and $\phi$ mesons
at small $x$ \cite{gk05, gk04}.

In this report, we  discuss the spin effects in the $\rho$ meson
leptoproduction. Within the MPA we calculate  the three $LL$, $TT$
and $TL$ transition amplitudes,  and afterwards cross sections and
SDME.

\section{Leptoproduction of  Vector Mesons in the GPD approach}
The leading-twist wave function $\hat \Psi^0_V$ describes the
longitudinally polarized vector mesons. To study the transversally
polarized light mesons, we use the higher order $k$- dependent
wave function $\hat \Psi^1_V$ proposed in \cite{koerner}
\begin{equation} \hat
\Psi_V= \hat \Psi^0_V+ \hat \Psi^1_V \nn.
\end{equation}
\begin{equation} \hat \Psi^0_V=
(\,/\hspace{-2.mm} V+m_V) \,/\hspace{-2.mm} \epsilon_V
         \phi_V(k,\tau)\nn.
\end{equation}
\begin{equation} \hat \Psi^1_V= [\frac{2}{M_V}\,
/\hspace{-2.2mm} V \,
                 /\hspace{-2.mm}
                  \epsilon_V /\hspace{-2.mm} K
         - \frac{2}{M_V} (\, /\hspace{-2.2mm} V - m_V) ( \epsilon_V \cdot K)]
         \phi'_V(k,\tau)\nn.
\end{equation}
Here  $\epsilon_V$ is a polarization of the vector meson with
momentum $V$, $\tau$ is the fraction of the meson momentum carried
by the quark, $M_V$ is the scale in the higher order wave function
which should be about the meson mass $m_V$. We use here
$M_V=m_V/2$

The gluon contribution to the leptoproduction amplitudes for $t
\sim 0$ and positive proton helicity reads as a convolution of the
hard subprocess amplitude  ${\cal H}^V$ and   a large distance
unpolarized  $ H^g$ and polarized  $\widetilde{H}^g$ gluon GPDs
(Fig.1):
\begin{eqnarray}\label{amptt-nf-ji}
  {\cal M}_{\mu'+,\mu +} &=& \frac{e}{2}\, {\cal C}_V\,
         \int_0^1 \frac{d\xb}
        {(\xb+\xi)
                  (\xb-\xi + i{\veps})}\nn\\
        &\times& \left\{\, \left[\,{\cal H}^V_{\mu'+,\mu +}\,
         + {\cal H}^V_{\mu'-,\mu -}\,\right]\,
                                   H^g(\xb,\xi,t) \right. \nn\\
       &+& \hspace{2.5mm}\left. \left[\,{\cal H}^V_{\mu'+,\mu +}\,
            -   {\cal H}^V_{\mu'-,\mu -}\,\right]\,
                        \widetilde{H}^g(\xb,\xi,t)\, \right\}\,.
\end{eqnarray}

Here $\mu$ ($\mu'$) denotes the helicity of the photon (meson),
$\xb$ is the  momentum fraction of the transversally polarized
gluons and the skewness $\xi$ is related to Bjorken-$x$ by
$\xi\simeq x/2$. The flavor factors are $C_{\rho}=1/\sqrt{2}$ and
${ C}_{\phi}=-1/3$.  The polarized $\widetilde{H}^g$ GPD at small
$\xb$ is much smaller than the unpolarized GPD $H^g$ and will be
important in the $A_{LL}$ asymmetry only.

 The subprocess amplitude ${\cal H}^V$ is represented as  the contraction of the hard
  part $F$, which is calculated perturbatively, and the
non-perturbative meson  wave function $ \phi_V$
\begin{equation}\label{hsaml}
  {\cal H}^V_{\mu'+,\mu +}\,\pm\,  {\cal H}^V_{\mu'-,\mu -}\,=
\,\frac{2\pi \als(\mu_R)}
           {\sqrt{2N_c}} \,\int_0^1 d\tau\,\int \frac{d^{\,2} \vk}{16\pi^3}
            \phi_{V}(\tau,k^2_\perp)\;
                  F_{\mu^\prime\mu}^\pm .
\end{equation}
  The wave function is chosen  in Gaussian form
\begin{equation}\label{wave-l}
  \phi_V(\vk,\tau)\,=\, 8\pi^2\sqrt{2N_c}\, f_V a^2_V
       \, \exp{\left[-a^2_V\, \frac{\vk^{\,2}}{\tau\bar{\tau}}\right]}\,.
\end{equation}
 Here $\bar{\tau}=1-\tau$, $f_V$ is the
decay coupling constant and the $a_V$ parameter determines the
value of average transverse momentum of the quark in the vector
meson. Generally, values of $f_V, a_V$ are  different for
longitudinal and transverse polarization of the meson.

The subprocess amplitude is calculated within the MPA
\cite{sterman} where we keep the $k^2_\perp$ terms in the
denominators of the amplitudes and the numerator of the TT
amplitude. The gluonic corrections are treated in the form of the
Sudakov factors which additionally suppress the end-point
integration regions. The model leads to the following hierarchy of
the amplitudes:
\begin{equation}
{\rm LL}:\;\;\;  {\cal M}_{0\,\nu,0\,\nu}^{V(g)} \propto
1\,;\qquad {\rm TL}:\;\;\;  {\cal M}_{0\,\nu,+\nu}^{V(g)} \propto
                                                  \frac{\sqrt{-t}}{Q};\qquad
{\rm TT}:\;\;\;{\cal M}_{+\nu,+\nu}^{V(g)} \propto \frac{\vk^2}{Q
M_V}.
 \end{equation}
Note, that the  $L \to T$ and $T \to -T$ transitions are small and
neglected in our analysis.

The GPD is a complicated function which depends on three
variables. For the small momentum transfer  $t \sim 0$ we  use the
double distribution form \cite{mus99}
\begin{equation}
  H^{g}(\xb,\xi,t) = \Big[\,\Theta(0\leq \xb\leq \xi)
    \int_{\frac{\xb-\xi}{1+\xi}}^{\frac{\xb+\xi}{1+\xi}}\, d\beta +
     \Theta(\xi\leq \xb\leq 1) \int_{\frac{\xb-\xi}{1-\xi}}
     ^{\frac{\xb+\xi}{1+\xi}}\, d\beta \,\Big]\,
 \frac{\beta}{\xi}\,f(\beta,\alpha=\frac{\xb-\beta}{\xi},t)
\end{equation}
with the simple factorizing ansatz for the double distributions
$f(\beta,\alpha,t)$
\begin{equation}
  f(\beta,\alpha,t\simeq 0) = g(\beta)\,\frac{3}{4}\,
\frac{[(1-|\beta|)^2-\alpha^2]}{(1-|\beta|)^{3}}.
\end{equation}

\begin{figure}[h]
\centering \mbox{\epsfysize=50mm\epsffile{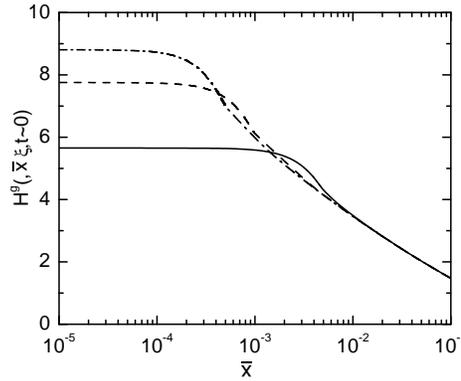}}
\caption{Model results for the GPD $H^g$ in the small $\xb$ range.
The solid (dashed,  dash-dotted) line represents the GPD at $\xi=
5\;(1\,,\; 0.5)\, \cdot 10^{-3}$}
\end{figure}
 In this
model, we calculate GPD \cite{gk05} through  the gluon
distribution $g(\beta)$ and take the CTEQ5M results~\cite{CTEQ}
for it . For the three $\xi$ values the gluon GPD is shown in Fig.
2.

\section{Cross section and spin observables}

The  $t$- dependence of the amplitudes is important in analyses of
experimental data. For simplicity we parameterize it in the
exponential form $M_{ii}(t)=M_{ii}(0)\; e^{t\,B_{ii}/2}$ for $
LL,\; TT,\; TL$ transitions. Experimentally, only the slope of the
$\gamma^* p\to Vp$ cross section is measured whereas information
about individual slopes $B_{ii}$  is absent. We suppose   that  $
B_{LL}\sim B_{TT} \sim B_{TL}$. Estimations for the $\rho$ meson
production amplitudes  are obtained using $f_{\rho L}=.216\gev$,
  $a_{\rho L}=0.522\gev^{-1}$; $f_{\rho T}=.170\gev$; $a_{\rho T}=0.65\gev^{-1}$.

\begin{figure}[h!]
\begin{center}
\begin{tabular}{cc}
\mbox{\epsfig{figure=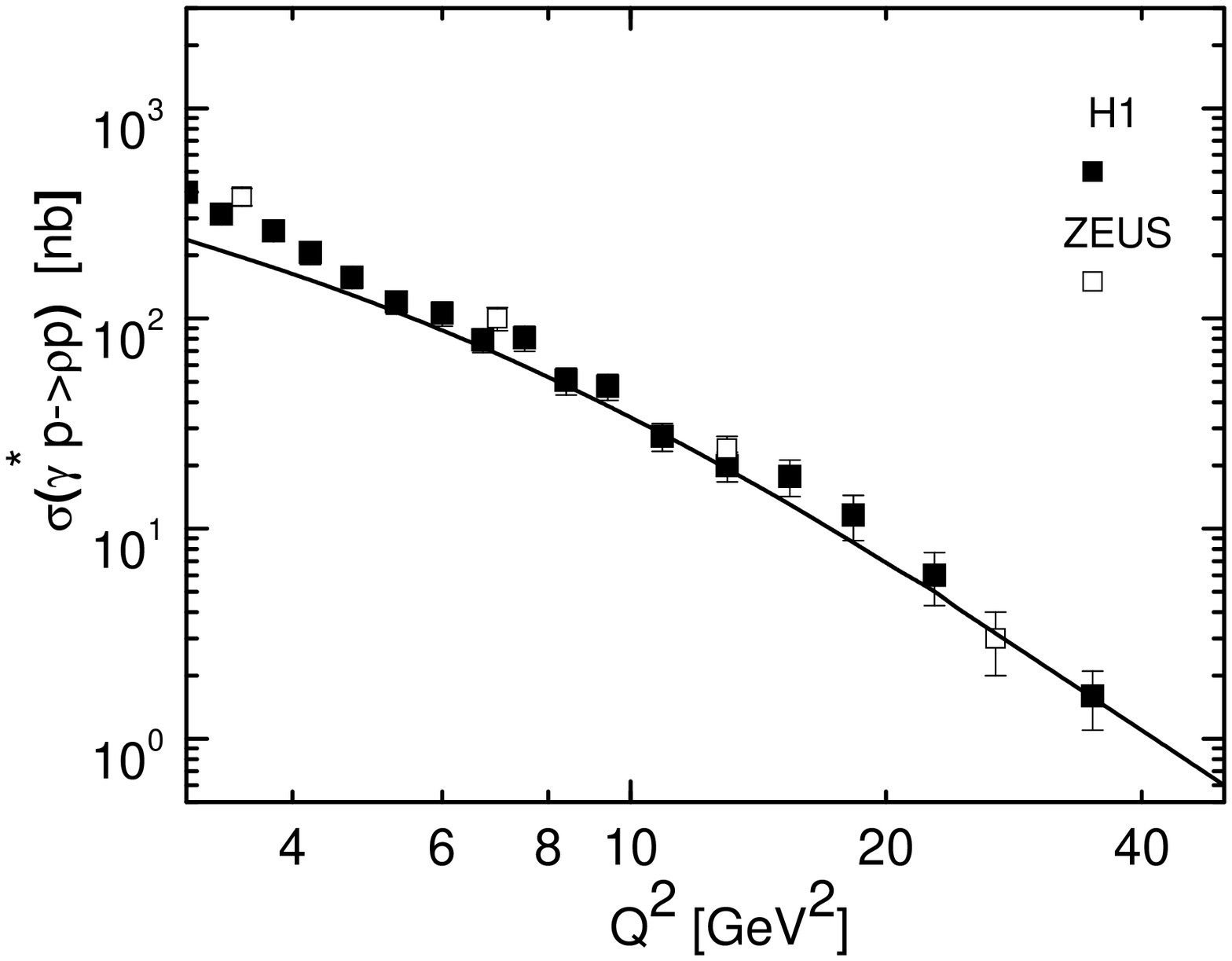,width=6.2cm,height=5.5cm}}&
\mbox{\epsfig{figure=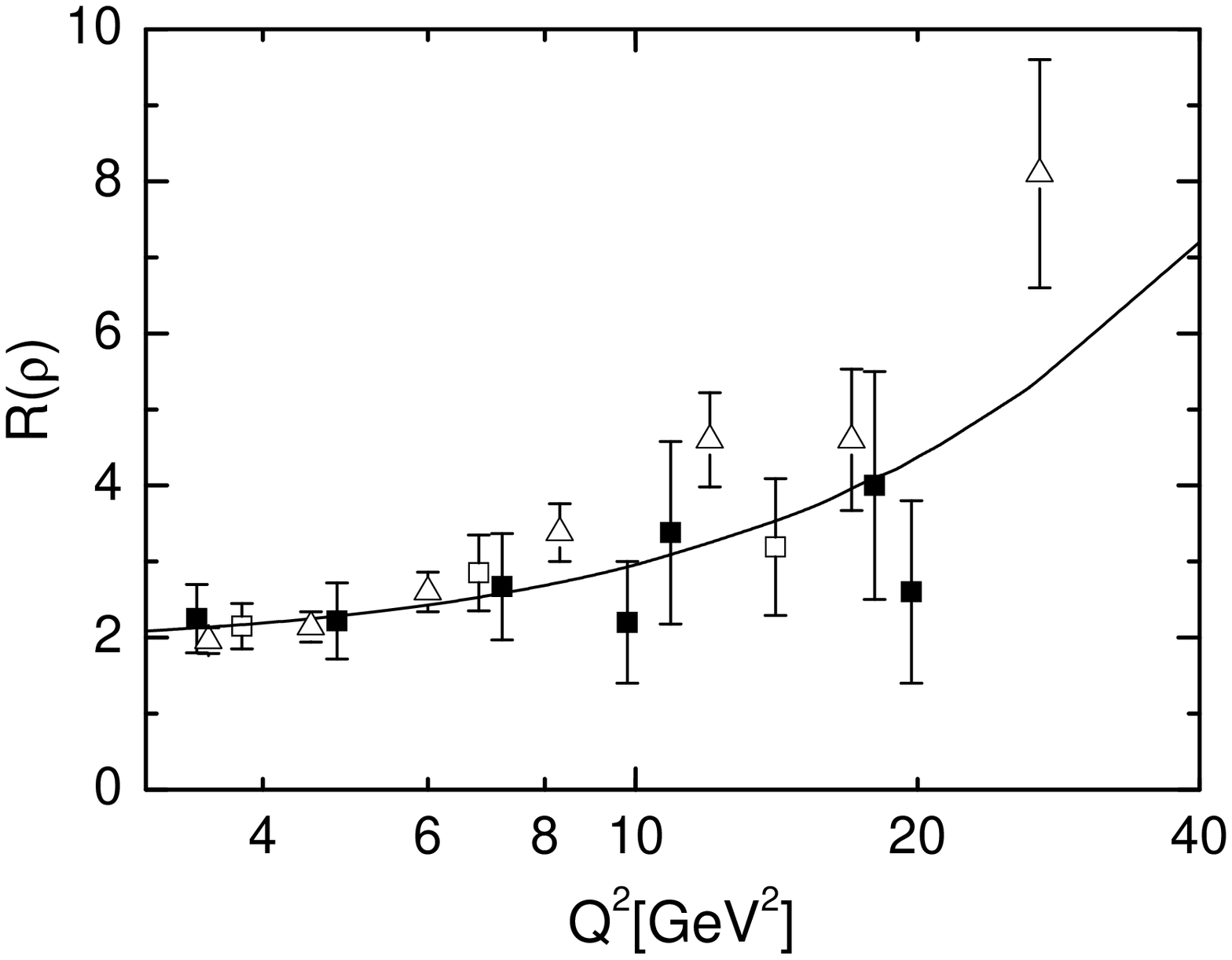,width=5.8cm,height=5.5cm}}\\
{\bf(a)}& {\bf(b)}
\end{tabular}
\end{center}
{\small{ Fig. 3a.} The cross section for $\gamma^*\, p\to \rho^0\,
p$ vs. $Q^2$ for $W=75\gev$. }\\
{\small{Fig. 3b.} The ratio of longitudinal and transverse cross
sections
 for the $\rho$  production vs. $Q^2$ at
 $W =75\, \gev$ and $t=-0.15\,\gev^2$. Data are  from H1 and ZEUS.}
\end{figure}

The cross section for the $\gamma^* p \to \rho p$
 production integrated over $t$  is shown
in Fig. 3a . Good agreement with  H1 and ZEUS  experiments
\cite{h1,zeus} at HERA is to be observed. Note that the
uncertainties in the gluon GPD provide about $20-30 \%$ errors in
the cross section.  The model results for the ratio of the cross
section with longitudinal and transverse photon polarization $R$
for the $\rho$ production are shown in Fig.3b. They are consistent
with H1 and ZEUS experiments  \cite{h1,zeus}.

\begin{figure}[h!]
\begin{center}
\begin{tabular}{ccc}
\mbox{\epsfig{figure=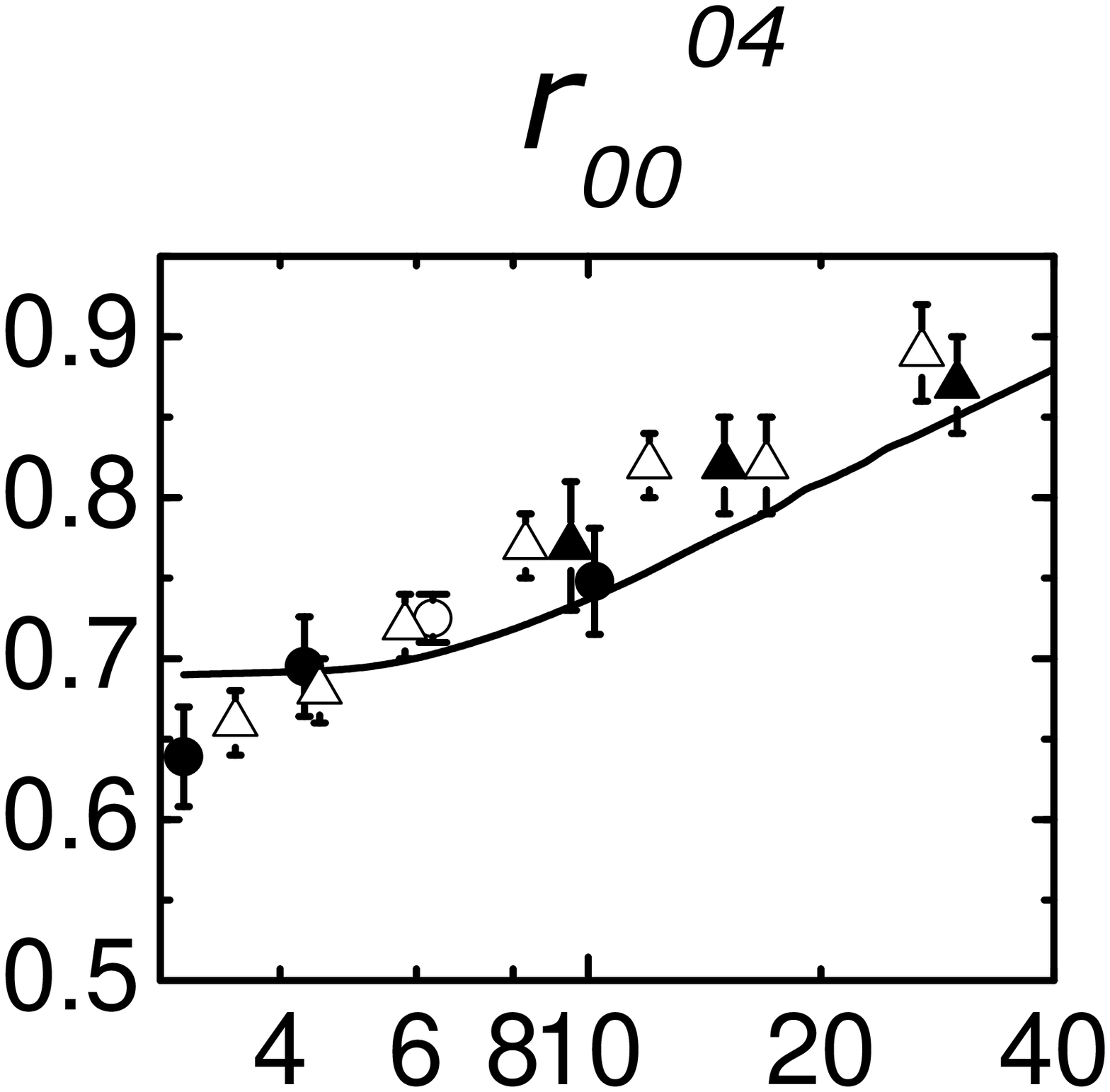,width=3.9cm,height=3.5cm}}&
\mbox{\epsfig{figure=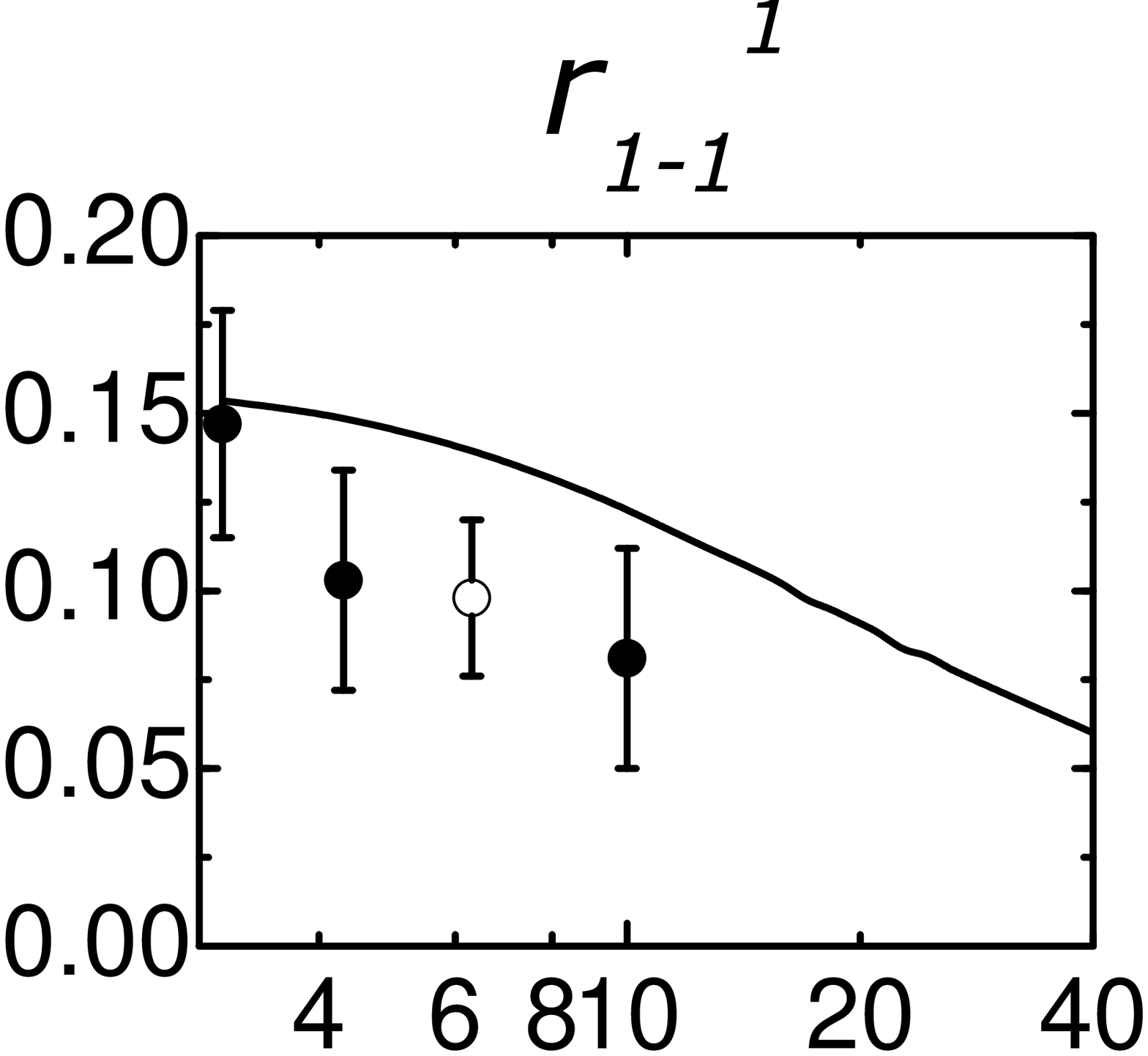,width=3.9cm,height=3.5cm}}&
\mbox{\epsfig{figure=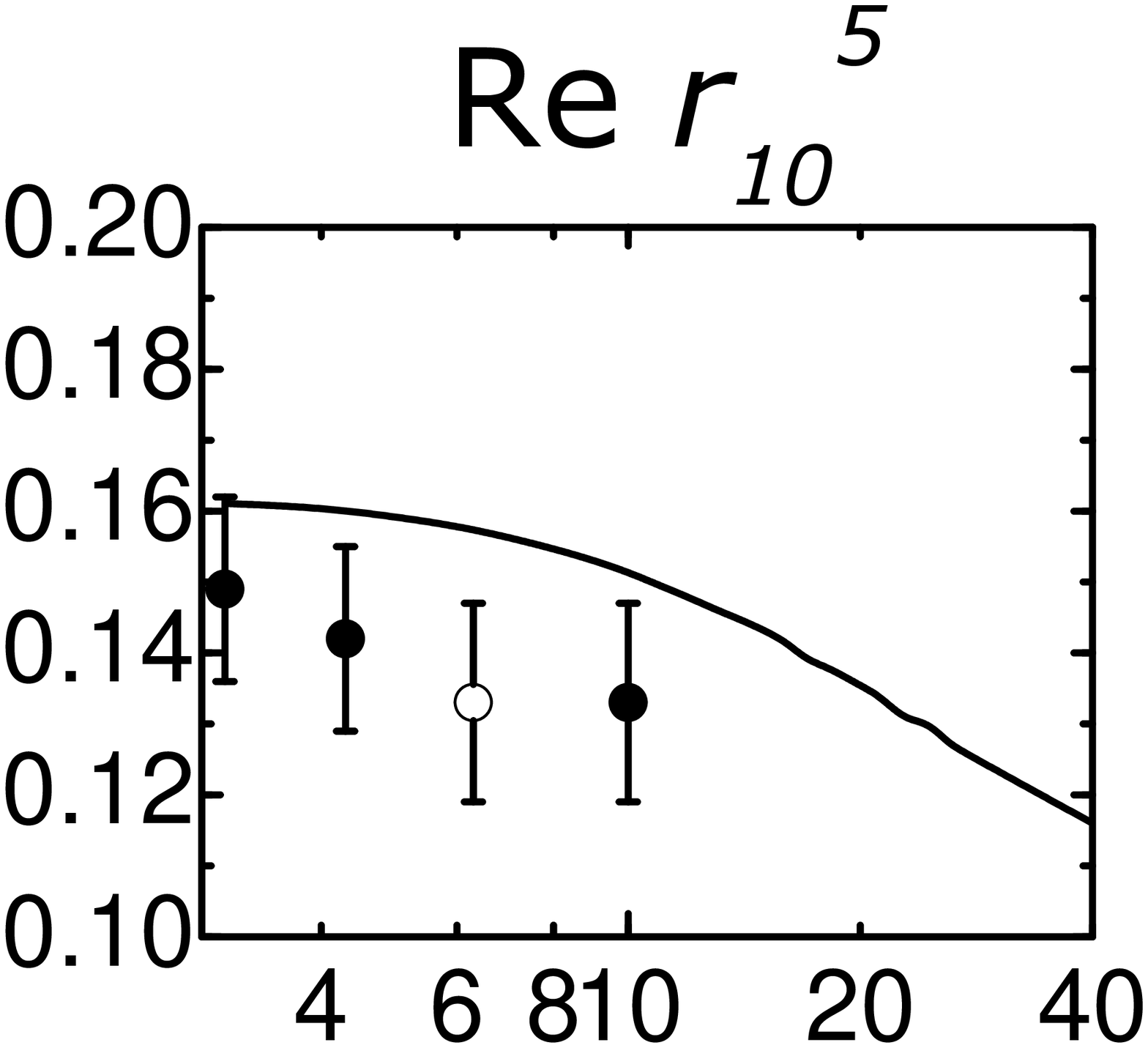,width=3.9cm,height=3.5cm}}\\
\end{tabular}
\phantom{aa}\vspace{-1mm} \centerline{$Q^2[\mbox{GeV}]^2$}
\end{center}
{\small{Fig. 4} The $Q^2$ dependence of SDME on the $\rho$
production at $t=-.15\gev^2$ and $W=75\gev$.   Data are taken from
H1 and ZEUS. }
\end{figure}

In Fig.4, we present our results for the three  SDME at DESY
energy range. Description of experimental data is reasonable. Our
results for other SDME can be found in \cite{gk05}

The $A_{LL}$ asymmetry for a longitudinally polarized beam and
target is sensitive to the polarized GPD.  Really, the leading
term in the $A_{LL}$ asymmetry integrated over the azimuthal angle
is determined through the interference between the unpolarized GPD
$H^g$ and the polarized $\widetilde{H}^g$ distributions.

We expect a small  value for the $A_{LL}$ asymmetry at high
energies because it is of the order of the ratio $\langle
\widetilde{H}^g\rangle/\langle H^g \rangle$ which is small  at low
$x$.
\begin{figure}
\begin{center}
\mbox{\epsfig{figure=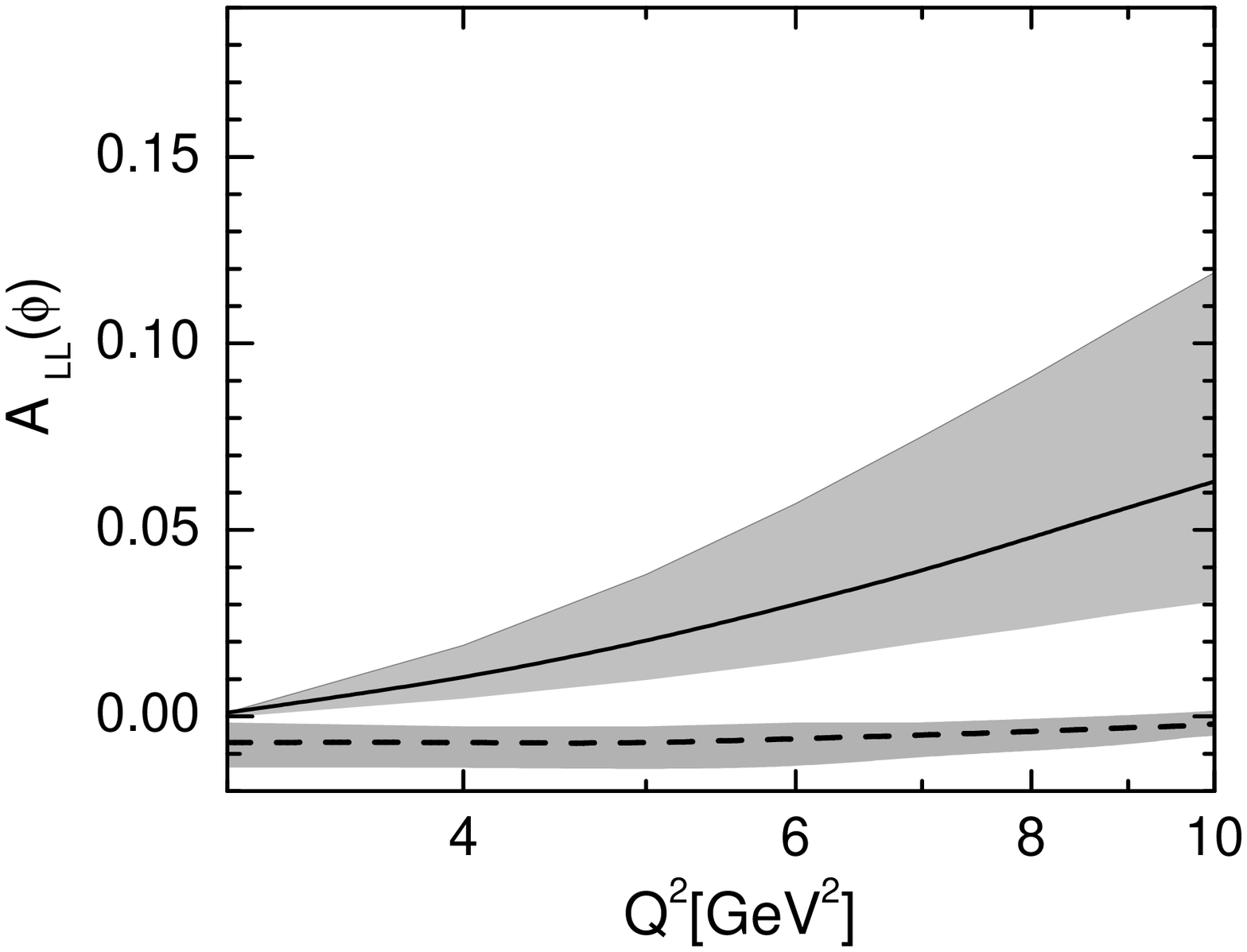,width=5.2cm,height=4cm}}
\end{center}
{\small{ Fig. 5.} Gluon contribution to the $A_{LL}$ asymmetry of
vector meson production at $W=5\,\gev$ (solid line) and
  $W=10\,\gev$ (dashed line); $y\simeq 0.6$. The shaded bands reflect
  the uncertainties in the $A_{LL}$ due to the  error in
  $\widetilde{H}^g \propto \Delta g$  }
  \label{yourname_fig4}
\end{figure}
At COMPASS energies $W=10 \mbox{GeV}$ the gluon contribution to
the $A_{LL}$ asymmetry of vector meson  production is  small
(Fig.5). At HERMES energies $W=5 \mbox{GeV}$ the major
contribution to the amplitudes comes from the region $0.1 \le \xb
\le 0.2$ where $\Delta g/g$ is not small, which leads to a large
value of the $A_{LL}$. The asymmetry is shown in Fig.5 together
with the uncertainties in our predictions due to the error in the
polarized gluon distribution. Note that our predictions should be
valid for the $\phi$ production. In the case of $\rho$ production
the polarized quark GPD should be essential in the $A_{LL}$
asymmetry in the low energy range.

 \section{Conclusion or Summary}

Light vector meson electroproduction at small $x$ was analyzed
here within the  GPD approach. The amplitudes were calculated
using the MPA with the wave function  dependent on the transverse
quark momentum. The transverse quark momentum   regularizes the
end-point singularities in the amplitudes with transversally
polarized photons. This gives a possibility to study spin effects
in vector meson production.  Within the GPD approach we find a
fine description of the $Q^2$ dependence of the cross section, $R$
ratio and   SDME for the $\rho$ meson production at HERA energies.
More results for $\rho$ and $\phi$ production can be found in
\cite{gk05}. Some comparison with the two-gluon exchange model
\cite{bro94}  can be found in \cite{gk05} too.

 We would like to point out that study of SDME gives  important
 information on different  $\gamma \to V$ hard amplitudes. By analyses of $t$
 dependencies of SDMEs  the diffraction peak slopes of
the $TT$ and $TL$ amplitudes can be defined. Unfortunately, the
data on spin observables have  large experimental errors. The new
experimental results with reduced errors in SDME are extremely
important. Information about polarized gluon distribution can be
obtained from the $A_{LL}$ asymmetry  on $\phi$ meson production
at HERMES and COMPASS energies  which is sensitive to $\Delta G$.

Thus, we can conclude that the vector meson photoproduction at
small $x$ is an excellent tool  to probe the gluon  GPD.

 \bigskip

 {\small This work is supported  in part by the Russian Foundation for
Basic Research, Grant 06-02-16215  and by the Heisenberg-Landau
program.}

 \end{document}